# Features of the configurations "small particle –plate" and "plate –plate" in the theory of fluctuation electromagnetic interaction

G.V.Dedkov and A.A.Kyasov

Kabardino –Balkarian State University, ul.Chernyshevskogo 173, Nalchik, 360004, Russia

e-mail: gv_dedkov@mail.ru

It is shown that the limiting transition from the geometrical configuration "plate –plate" to configuration "small particle –plate" being frequently used in the theory of Lifshitz –Pitaevskii, is not continually true. On the other hand, the known solution to the problem in the last configuration can be used to verify the generalizations of the theory being worked out in the former configuration.

In recent time, interest in fluctuation electromagnetic interaction (FEI) between condensed bodies has become considerably greater (see reviews [1-7]). Traditionally, within the scope of FEI one considers the conservative van –der –Waals and Casimir forces, radiative heat exchange and dissipative forces arising at relative motion of the bodies divided by a vacuum gap. The study of FEI between moving bodies, in contrast to the static case, reveals many intriguing features.

Historically, since nearly creation of the theory of electromagnetic fluctuations by S.M.Rytov [8], it has been applied by E.M.Lifshitz [9] in calculation of the interaction force between two thick plates at rest (configuration 1, fig.a). From the very beginning, configuration "small particle –plate" (configuration 2, fig.b) has attracted much less attention (for more details see [2,7]), because the force of attraction of the particle to the plate could be obtained using the limit of rarified medium for the substance of one of the plates: $\varepsilon(\omega) - 1 = 4\pi n \alpha(\omega) \to 0$ [9-11], $n$ and $\alpha(\omega)$ are the corresponding volume atomic density and polarizability of the rarified material, $\varepsilon(\omega)$ is the dielectric permittivity. This has led to an opinion that configuration 2 is only a special case of configuration 1 even in the case out of equilibrium at a definite temperature difference between the bodies, or under their relative motion [5,12]. Of principle self–dependent importance of configuration 2 stems from our works [7] (see also references), in which an exact relativistic solution was obtained for the dynamic problem 2 at arbitrary temperature of the particle ($T_1$), surrounding vacuum background ($T_2$), and different material properties of these bodies.

This work aims at demonstrating two examples of inadequacy of the transition «1 → 2». On the other hand, we draw attention to a possibility of using the yet obtained exact results relevant to configuration 2 in an "opposite direction": to testing the theories under development in configuration 1 in dynamic and out of equilibrium situations.

Let us consider conservative fluctuation electromagnetic interaction between a small particle (neutral atom in the ground state) and a thick plate at $T_1 = T_2 = 0$, $V = 0$. From the general expression for the force $F_z$, obtained in [7], it follows

$$F_z = -\frac{\hbar}{\pi}\int_0^\infty d\xi \int_0^\infty dk\, k \exp(-2\sqrt{k^2 + \xi^2/c^2}\, z)[R_e(i\xi,k)\alpha_e(i\xi) + R_m(i\xi,k)\alpha_m(i\xi)] \quad (1)$$

$$R_e(\omega,k) = (2k^2 - \omega^2/c^2)\Delta_e + (\omega/c)^2 \Delta_m \quad (2)$$

$$R_m(\omega,k) = (2k^2 - \omega^2/c^2)\Delta_m + (\omega/c)^2 \Delta_e \quad (3)$$

where $\Delta_e$ and $\Delta_m$ are the Fresnel reflection amplitudes for the electromagnetic waves having $P-$ and $S-$polarization, $\alpha_e$, $\alpha_m$ are the dipole dielectric and magnetic polarizabilities of the particle, other quantities have their conventional meaning. With no account of magnetic polarization of the particle, Eq.(1) exactly coincides with that one obtained using the limiting transition «1 → 2» [13]. Thus, Eq.(1) describes the Casimir–Polder force applied to the ground state atom ($T_1 = 0$) near a cold wall. If the ground state atom or a cold nanoparticle interacts with the heated wall ($T_2 = T \neq 0$), then, as follows from Eq. (54) in [7] (see aforesaid Ref.),

$$F_z = -2k_B T \sum_{n=0}^\infty a_n \int_0^\infty dk\, k [R_e(i\xi_n,u)\alpha_e(i\xi_n) + R_m(i\xi_n,u)\alpha_m(i\xi_n)]\exp\left(-2\sqrt{k^2 + \xi_n^2/c^2}\, z\right) + \Delta F_z \quad (4)$$

$$\Delta F_z = \frac{2\hbar}{\pi} \int_0^\infty d\omega \, \Pi(\omega,T) \alpha_e''(\omega) \, \text{Re} \left\{ \int_0^\infty dk k \left[ \exp(-2q_0 z) R_e(\omega,k) \right] \right\} + \{ \alpha_e'' \to \alpha_m'', R_e \to R_m \} \quad (5)$$

where $a_n = 0.5$ at $n=0$ and $a_n = 1$ at $n \neq 0$, $\Pi(\omega,T) = (\exp(\hbar\omega/k_B T) - 1)^{-1}$ denotes the Planck's factor, $q_0 = (k^2 - \omega^2/c^2)^{1/2}$, and double primed quantities correspond to imagine components.

At thermal equilibrium $T_1 = T_2 = T$, the additional term $\Delta F_z$ is absent and both methods of calculation lead to the same result [7,14], but out of equilibrium the conventional transition «1→2», in contrast to (4), allows to retrieve only the first term [14]. Therefore, the involved transition proves to be satisfactory only when the temperature of the plate equals the temperature of constituent particles, which it is consist of. Under a subdivision of the plate on small nanoparticles, the temperature of each of them being the ordinary thermodynamic temperature, whereas under a subdivision on separate atoms one must speak about the temperature of its inner states of freedom, which determine fluctuations of the corresponding multipolar moments (as long as the temperature conception is valid).

As far as dynamic configurations 1 and 2 is concerned, when the dissipative force appears side by side with the conservative attraction force, one can formulate the following general requirement: correct formulae for the conservative –dissipative forces obtained in dynamic configuration 1 must lead to the corresponding ones in dynamic configuration 2 using transition «1→2». Let us employ this principle to examine the theory of vacuum friction, developed in [5,6] (see also references). For simplicity, we restrict this discussion by linear velocity approximation. Thus, using the limiting transition «1→2», the authors of [5,6] in the case $T_1 = T_2 = T$ obtained the following expression for the vacuum friction force applied to a small particle near the wall (Eq.(92) in [5], in our notations)

$$F_x = -\frac{\hbar V}{\pi} \int_0^\infty d\omega (-\partial \Pi / \partial \omega) \alpha_e''(\omega) \int_0^\infty dk k^3 \, \text{Im} \left[ \frac{\exp(-2kz)}{k} \left( 2k^2 \Delta_e + (\omega/c)^2 \Delta_m \right) \right] \quad (6)$$

In contrast with that, as it follows from our general solution obtained in configuration 2 [7] with no account of magnetic polarization of the particle and contribution from vacuum Planck's modes being independent of distance $z$:

$$F_x = -\frac{\hbar V}{\pi} \int_0^\infty d\omega (-\partial \Pi/\partial \omega) \alpha_e''(\omega) \left[ \int_{\omega/c}^\infty dk k^3 \text{Im}\left( \frac{\exp(-2q_0 z)}{q_0} R_e(\omega,k) \right) + \int_0^{\omega/c} dk k^3 \text{Im}\left( \frac{\exp(2i\tilde{q}_0 z)}{\tilde{q}_0} \tilde{R}_e(\omega,k) \right) \right] \quad (7)$$

where $\tilde{q}_0 = (\omega^2/c^2 - k^2)^{1/2}$, $\tilde{R}_e(\omega,k)$ is described by (2) with the replacements $\Delta_e \to \tilde{\Delta}_e, \Delta_m \to \tilde{\Delta}_m$, and complete formulae for $\Delta_{e,m}$ and $\tilde{\Delta}_{e,m}$ are given in [7]. In the nonrelativistic limit ($c \to \infty$) Eqs. (6),(7) are reduced to identical result [15] ($\varepsilon(\omega)$ is the dielectric permittivity of the plate material)

$$F_x = -\frac{3\hbar V}{2\pi z^5} \int_0^\infty d\omega (-\partial \Pi/\partial \omega) \alpha_e''(\omega) \text{Im}\left( \frac{\varepsilon(\omega)-1}{\varepsilon(\omega)+1} \right) \quad (8)$$

However, in the relativistic case with account of retardation, the difference between (6) and (7) becomes of principal importance. For instance, a very characteristic is the difference between the expressions $2k^2 \Delta_e + (\omega/c)^2 \Delta_m$ and $R_e(\omega,k)$ (see Eq. (2)), because the factor $2k^2 \Delta_e + (\omega/c)^2 \Delta_m$ appears in many papers of the authors [5,6]. By the way, just the same factor as in Eq.(2) arises when solving the boundary electrodynamic problem for the Maxwell equations with point dipole fluctuating sources [2,7]. That is in accordance with calculation of the conservative interaction forces between an atom and the plate [14,16]. On the other hand, as was shown in [17,18], just the same factor determines spectral density of equilibrium electromagnetic field near the heated surface. Particularly, for example, one has obtained the following expression for the electric component of the field [18] (for the magnetic component one needs simply to change $\Delta_e$ by $\Delta_m$):

$$(\mathbf{E}^2)_{\omega k} = 2\pi \hbar \coth(\hbar \omega/k_B T) \text{Im}\left( \frac{\exp(-2q_0 z)}{q_0} \left[ (2k^2 - \omega^2/c^2)\Delta_e + (\omega/c)^2 \Delta_m \right] \right) \quad (9)$$

Eq.(9) results from the same Green function representation for the fluctuation electromagnetic field, which is used in derivation of Eqs.(1), (4) in the case of thermal equilibrium. Therefore, Eq.(6) proves to be in contradiction with its own ground, because Eqs. (1) and (4) (without the second term) are the well recognized classical results. Moreover, in error is Eq. (58) in [5] for the particle heating rate.

As a curious instance, we give one more expression for the friction force in configuration 2, which has been obtained by the same authors in Ref. [19]:

$$F_x = -\frac{\hbar V}{\pi}\int_0^\infty d\omega\,\omega(-\partial\Pi/\partial\omega)\int_0^\infty dk\,\exp(-2kz)\alpha''(\omega)\left[2(2k^2+\omega^2/c^2)\Delta_e'' + (\omega/c)^2\Delta_m''\right] \qquad (10)$$

Eq.(10), contrary to the author's claim, despite having been obtained in the framework of Lifshitz theory using the limiting transition «1 → 2», does not result in true nonrelativistic limit (8) and has incorrect dimension of force.

Therefore, the theory of vacuum friction in configuration 1, being reproduced with some modifications in a series of works of the authors [5,6,19], proves to be inconsistent. Thus, the dynamic generalization of the Lifshitz theory still remains a very important unresolved problem.

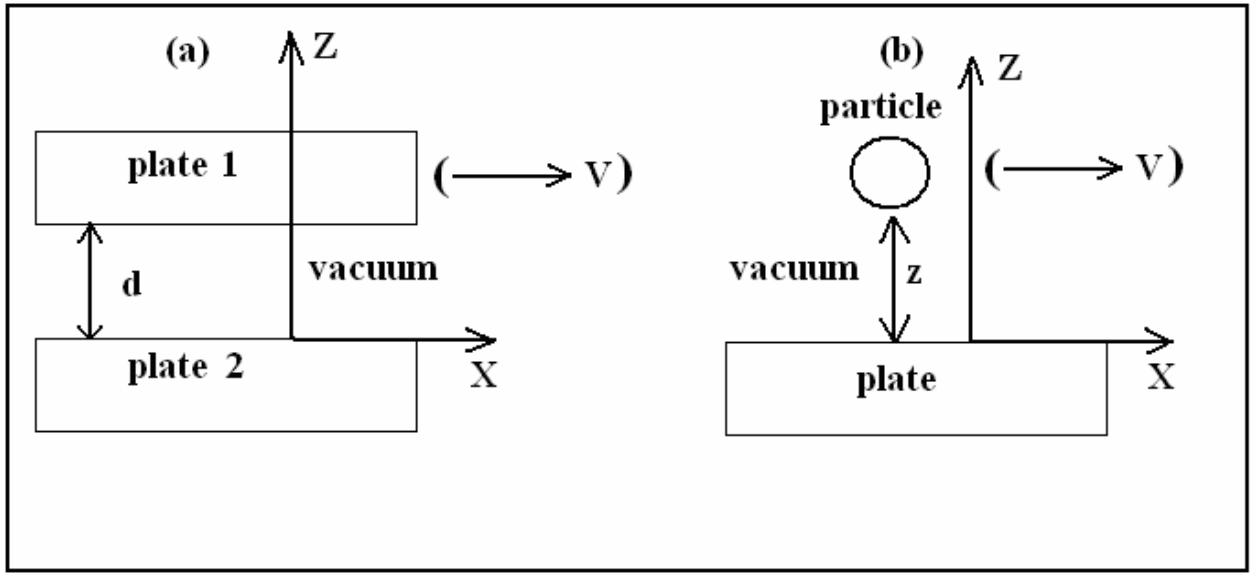

Configuration "plate –plate" (a) and "small particle –plate" (b). In the case (b) $R << z$ ($R$ is the particle radius). In the static case $V = 0$. In context of this paper, both configurations correspond to the thermal equilibrium of all the bodies (and vacuum) at temperature $T$.